\documentclass[showkeys,showpacs,longbibliography,prb]{revtex4-2}
\usepackage{amsmath}
\usepackage{color}
\usepackage{graphicx}
\usepackage{ulem}

\newcommand{\VEC}[1]{{\boldsymbol{ #1}}}

\newcommand{\Fig}{{Fig.}}
\newcommand{\Eq}{{Eq.}}
\newcommand{\bef}{BeF$_2$}
\newcommand{\znf}{ZnF$_2$}
\newcommand{\gru}{Gr\"uneisen}
\newcommand{\PThreeOne}{$P3_121$}
\newcommand{\PFourOne}{$P4_12_12$}
\newcommand{\PFourThree}{$P4_32_12$}
\newcommand{\Ctwoc}{$C12/c1$}
\newcommand{\rutile}{$P4_2/mnm$}
\newcommand{\pbesol}{PBE$\_$sol}
\newcommand{\invcm}{cm$^{-1}$}
\newcommand{\alphaV}{\alpha_v}

\begin{document}
\title{A first-principles investigation of the linear thermal expansion coefficients of \bef{}: Giant thermal expansion}

\author{Chee Kwan Gan}
\email{ganck@ihpc.a-star.edu.sg}
\affiliation{Institute of High Performance Computing, 1 Fusionopolis Way, \#16-16 Connexis 138632, Singapore}
\author{Abdullah I. Al-Sharif, Ammar Al-Shorman}
\affiliation{Department of Physics, Yarmouk University, Irbid-21163, Jordan}
\author{Abdallah Qteish}
\email{aqteish@yu.edu.jo}
\affiliation{Department of Physics, Yarmouk University, Irbid-21163, Jordan}

\date{\today }

\begin{abstract}
We present the results of a theoretical investigation of the
linear thermal expansion coefficients (TECs) of \bef{}, within a
direct \gru{} formalism where symmetry-preserving deformations are
employed. The required physical quantities such as the optimized
crystal structures, elastic constants, mode \gru{} parameters, and
phonon density of states are calculated from first-principles. \bef{}
shows an extensive polymorphism at low pressures, and the lowest energy
phases [$\alpha$-cristobalite with space group (SG) \PFourOne{} and
its similar phase with SG \PFourThree{}] are considered in addition to
the experimentally observed $\alpha$-quartz phase. For benchmarking
purposes, similar calculations are performed for the rutile phase of \znf{},
where the volumetric TEC ($\alphaV$), derived from the
calculated linear TECs along the $a$ ($\alpha_a$) and $c$ ($\alpha_c$)
directions, is in very good agreement with experimental data and previous
theoretical results. For the considered phases of \bef{}, we do not find
any negative thermal expansion (NTE). However we observe 
diverse thermal properties
for the distinct phases. The linear TECs are
very large, especially $\alpha_c$ of the $\alpha$-cristobalite phase
and its similar phase, leading to giant $\alphaV$ ($\sim 175 \times
10^{-6}$~K$^{-1}$ at 300~K). The giant $\alphaV$ arises from large
\gru{} parameters of low-frequency phonon modes,
and the $C_{13}$ elastic constant that is negatively signed and large in magnitude 
for the $\alpha$-cristobalite phase. The
elastic constants, high-frequency dielectric constants, Born effective
charge tensors, and thermal properties of the above phases of \bef{}
are reported for the first time and hence serve as predictions.
\end{abstract}


\maketitle
\section{Introduction}

Beryllium fluoride (\bef) is known to exist in glass and crystalline phases and
has a variety of technological applications. \bef{} glass has a large bandgap of
about $13.8$~eV, the lowest refractive index and highest Abbe number of any
inorganic material, and exceptional resistance to damage. These properties
have enabled the manufacturing of special glasses (from \bef{} and its mixtures
with fluorides and other difluorides) that have excellent transmittance in the UV
region\cite{Parker89,Gan95v184} and for use in high-power laser systems\cite{Weber78v32}. The LiF–\bef{} mixture is
a primary coolant and fuel solvent in molten salt nuclear reactors\cite{Mei13v50}. In protein
crystallography, \bef{} is used to restrict protein motion to facilitate the
crystallography process\cite{Kagawa04v23}. Very recently, crystalline \bef{} is predicted to be a
better neutron filter than MgF$_2$, which has been considered an effective neutron
filter candidate\cite{Qasir17v50}. 
The main aim of this work is to investigate, for the first time, the linear thermal expansion coefficients (TECs) of
a few low-energy crystalline phases of \bef{}.
We also consider a benchmark system \znf{} that
has exceptional electric and optical properties, and interesting technological applications ranging 
from catalysis to spectroscopy and laser applications\cite{Kaawar15v1653}.

Single crystal \bef{} has been grown and found to have a crystal structure
remarkably similar to that of the $\alpha$-quartz (SiO$_2$) structure\cite{Ghalsasi11v50}, which has a
trigonal symmetry with space group (SG) \PThreeOne{} (\#152). A recent first-principles study has
revealed that \bef{} shows extensive polymorphism at low pressures\cite{Nelson17v95}. 
Interestingly, three crystal phases [namely, (i) the $\alpha$-cristobalite phase that has a tetragonal
symmetry with SG \PFourOne{} (\#92), (ii) a similar phase to the $\alpha$-cristobalite phase (hereafter referred to as the $\alpha^{\prime}$-cristobalite phase) 
with SG \PFourThree{} (\#96) , and (iii) the $C2/c$-$2\times 4$ phase with
SG \Ctwoc{} (\#15)] are predicted to be energetically more stable than $\alpha$-quartz. 
However, these phases have a very small stability pressure range (less than $0.4$~GPa),
and the $\alpha$-quartz phase transforms to the coesite-I phase SG C2/c at $3.1$~GPa. The high-pressure phases of \bef{} have been the subject of other
first-principles calculations\cite{Rakitin15v17}. Very recently, first-principles calculations have also
been employed to construct the $P$-$T$ phase diagram of \bef{}\cite{Masoumi21}. The HSE06
optical bandgap of the $\alpha$-quartz structure is found to be about $10.6$~eV, and 
increases by increasing the applied pressure\cite{Nelson17v95}. The lattice vibrations, 
inelastic scattering cross-sections, and neutron transmission of \bef{} have been thoroughly
investigated\cite{Qasir17v50} using first-principles calculations and compared to those of
MgF$_2$.

The benchmark system \znf{} crystallizes in the tetragonal rutile structure with SG \rutile{} (\#136).
Very recently, Raman scattering measurements with the use of the diamond anvil cell have been employed to investigate the structural phase transformations
of \znf{} under high pressures\cite{Kurzydlowski20v59}. This experimental work is supplemented by first-principles calculations. 
In addition to the structural stability and pressure variation of the Raman active phonon modes, the electronic bandgap of the
considered phases as a function of pressure has been investigated at the HSE06 level. Neutron diffraction has been employed to study the temperature dependence of the lattice parameters and unit cell volume of \znf{}\cite{Chatterji11v98},  and NTE has been observed in a small temperature range (below 75 K). This NTE behavior has been supported by first-principles calculations\cite{Chatterji11v98,Wang12v47}. 
However,  only the volumetric TEC has been theoretically investigated.

In the present work, the linear TECs of \bef{} and \znf{} are 
investigated by employing the recently introduced direct approach\cite{Gan19v31} in which the 
symmetry of the deformed structures could be preserved. 
Since this approach has not been applied to systems with tetragonal symmetry, \znf{} is thus chosen
as a suitable benchmark system because of its tetragonal crystal structure, in addition to the existence of experimental and previous theoretical results of
its volumetric thermal expansion.
The elastic constants and phonon frequencies required to compute linear TECs are calculated from first-principles. For \bef{}, the  $\alpha$-quartz, $\alpha$-cristobalite and $\alpha^{\prime}$-cristobalite
phases are considered. Moreover, the relative stability of the above three
phases of \bef{} will be also investigated using different levels of approximation of
the exchange-correlation potential.

\section{Methodology}
\label{sec:method}

The linear TECs of the considered phases of \znf{} and \bef{} are calculated within the \gru{} formalism following the procedure described 
in Refs.~[\onlinecite{Gruneisen26v10,Barron80v29,Schelling03v68,Ding15v5,Gan15v92,Liu17v121,Liu18v154}].
To compute the mode \gru{} parameters we considered two types of symmetry-preserving deformations obtained by changing the in-plane ($a$) 
or out-of-plane ($c$) lattice parameters by $\pm 0.5\%$. These deformations allow for the full utilization of the tetragonal or trigonal point-group 
symmetry\cite{Gan21v259} of the considered systems, which minimizes the required number of independent atomic displacements (i.e., number 
of supercells) to calculate the phonon frequencies within the direct method\cite{Frank95v74,Parlinski97v78,Kresse95v32,Gan10v49,Liu14v16}. The
amplitude of atomic displacements, from the corresponding equilibrium positions, is $0.015$~\AA. The supercell sizes are of 
$2\times 2 \times 3$ for \znf{}, and $2 \times 2 \times 2$ for the $\alpha$-quartz, $\alpha$-cristobalite, and $\alpha^{\prime}$-cristobalite phases of \bef{}. The adequacy of these 
supercells have been checked by considering larger ones for each of these systems, and we found that such 
actions do not alter appreciably our main results and conclusions. The determination of the linear TECs also requires the 
elastic constants that may be obtained through fittings of energy versus strain curves\cite{Dalcorso16v28,Gan18v151}. Specifically, these TECs at temperature $T$ in the $a$ ($\alpha_a$) and $c$ ($\alpha_c$) directions of the above systems are given by 
\begin{equation}
\left[
\begin{matrix}
       \alpha_a(T)  \\
       \alpha_c(T) 
  \end{matrix}
\right]
= \frac{1}{\Omega D}
\left[
    \begin{matrix}
    C_{33} & -C_{13} \\
    -2C_{13} & (C_{11} + C_{12} )
\end{matrix}
\right]
\left[
\begin{matrix}
I_a(T) \\ I_c(T)
\end{matrix}
\right]
\label{eq:LTEC}
\end{equation}
where $C_{ij}$ are the elastic constants, $D = (C_{11} + C_{12}) C_{33} - 2C_{13}^2$ and $\Omega$ is 
the volume of the primitive cell. The phonon density of states (PDOS) weighted by the \gru{} parameters are 
\begin{equation} 
\Gamma_i(\nu) = \frac{\Omega}{(2\pi)^3} \sum_{\lambda} \int_{\rm BZ}  d\VEC{q}\  \delta( \nu - \nu_{\lambda\VEC{q}}) \gamma_{i,\lambda\VEC{q}}, 
\label{eq:Gammanu} 
\end{equation} 
with the \gru{} parameter $\gamma_{i,\lambda\VEC{q}} = -\nu_{\lambda\VEC{q}}^{-1} \partial \nu_{\lambda\VEC{q}}/\partial \epsilon_i$ for the deformation of type $i$ (with $i=a$ for in-plane and $i=c$ for out-of-plane deformations). 
The derivative $\partial \nu_{\lambda\VEC{q}}/\partial \epsilon_i$ measures the change of the frequency $ \nu_{\lambda\VEC{q}} $  with
respect to the strain parameter $\epsilon_i$\cite{Gan19v31}.
The summation is over all frequencies 
$\nu_{\lambda\VEC{q}}$ for the phonon band index $\lambda$ and $\VEC{q}$ vector in the Brillouin zone (BZ). The $I_i(T)$ 
are calculated from
\begin{equation} 
I_i(T) = \int_{\nu_{\rm min}}^{\nu_{\rm max}} d\nu\
\Gamma_i(\nu)c(\nu,T),
\end{equation}
where $c(\nu,T) = k_B (r /\sinh r)^2$ is the contribution of the phonon modes with frequency $\nu$ to the specific heat. Here, 
$r = h\nu/2k_B T$, and $h$ and $k_B$ are the Planck and Boltzmann constants, respectively. The maximum (minimum) frequency is denoted by $\nu_{\rm max}$ ($\nu_{\rm min}$).

The DFT calculations of the optimized structural parameters, phonon frequencies, and elastic constants are performed by employing 
the projector augmented wave (PAW) method, as implemented in the Vienna Ab-Initio Simulation Package (VASP). A relatively high cutoff 
energy of $600$~eV is used for the plane-wave basis. Geometry optimization is stopped when the maximum force on each atom is less than $10^{-3}$~eV/\AA{}. We find that phonon frequency shifts are more consistent when we use the local density approximation (LDA) 
for \bef{}, and the \pbesol{} functional of the generalized gradient approximation (GGA) for \znf{}. Therefore, for the 
linear TECs only the results of these calculations are reported. 

\begingroup
\squeezetable
\begin{table}
\caption {\label{tab:struct} 
Calculated lattice constants and internal parameters of the $\alpha$-quartz 
and $\alpha$-cristobalite phases of \bef{}. Also shown are the relative energy ($\Delta E$) of the 
$\alpha$-cristobalite phase with respect to that of $\alpha$-quartz, and the available experimental data 
(measured at 100 K) and other theoretical results.}
\begin{ruledtabular}
\begin{tabular}{lllllllllc}
Phase & Approach  & \multicolumn{2}{c}{Lattice constants} & & \multicolumn{4}{c}{Internal parameters} & $\Delta E$ (meV) \\
\cline{3-4} \cline{5-9} 
 &  & $a$ (\AA)  & $c$ (\AA) & & $x_1$ & $x_2$ & $y_2$ & $z_2$             &  \\  \hline
 $\alpha$-quartz &  LDA       & 4.5958 & 5.0529 & & 0.4579 & 0.4098 & 0.2867 &  0.2290  &        \\
          &  \pbesol{}  & 4.7301 & 5.1814 & & 0.4662 & 0.4138 & 0.2737 & 0.2188 &        \\   
          &  PBE       & 4.8497 & 5.3070 & & 0.4756 & 0.4176 & 0.2575 & 0.2053 & \\   
          &  PBE\cite{Qasir17v50} & 4.8282 & 5.2837 & & 0.4740 & 0.4171 &  0.2601 & 0.2075 & \\
          &  LDA\cite{Yu13v169}    & 4.6663 & 5.1608 & &        &        &         &        & \\   
          &  Expt.\cite{Ghalsasi11v50} & 4.7390  & 5.1875   && 0.4700 & 0.4164  & 0.2671 & 0.2131 & \\ \hline 
$\alpha$-cristobalite &  LDA      & 4.5967 & 6.1773                && 0.3226 & 0.2230 &  0.1454 &  0.2000 &  25    \\
          &  \pbesol{} & 4.8087 & 6.5984              && 0.3044 & 0.2378 & 0.1112 & 0.1825 &   $-2$  \\
          &  PBE       & 4.8934 & 6.7428              && 0.2988 & 0.2400 & 0.1001 & 0.1769 &   $-9$  \\  
          &  LDA\cite{Yu13v169}      & 4.695 & 6.318 &&        &        &        &        &    \\ 
          &  LDA\cite{Masoumi21} & 4.684 & 6.373 &&        &        &        &        &    \\  
          &  PBE\cite{Masoumi21} & 4.960 & 6.910 &&        &        &        &        &   \\
\end{tabular}
\end{ruledtabular}
\begin{tabbing}
\end{tabbing}
\end{table}
\endgroup

\begin{figure}[htbp]
    \begin{center}
    \includegraphics[scale = 0.40]{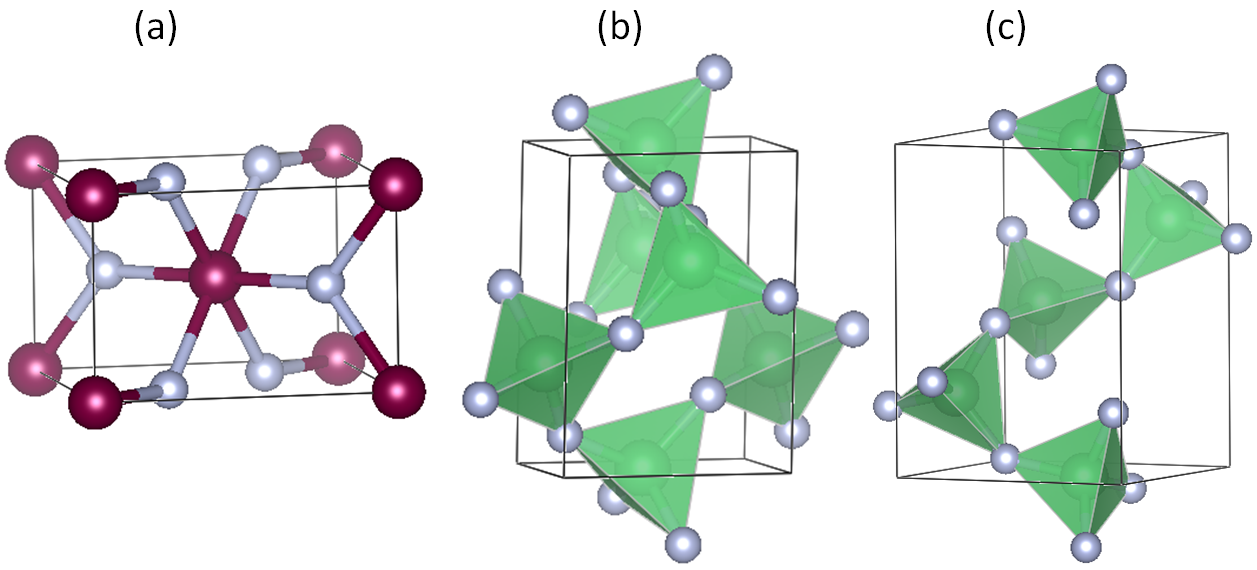} \\ 
    \caption{The crystal structures of (a) rutile \znf{},  (b) $\alpha$-quartz \bef{},  and (c) $\alpha$-cristobalite \bef{}. The 
    $\alpha^{\prime}$-cristobalite \bef{} structure which is similar to the $\alpha$-cristobalite \bef{} structure is not shown. 
The small gray balls represent the F atoms. The $c$ axis of the three crystal structures is along the vertical direction.}
    \label{fig:structures}
       \end{center}
\end{figure}

\section{Results and discussion}
\label{sec:results}

\subsection{Structural properties}

We start with the benchmark system, \znf{}, which at ambient conditions crystallizes in the rutile structure, shown in \Fig~{\ref{fig:structures}(a). This crystal 
structure has a tetragonal symmetry and six atoms per primitive unit cell. The two Zn atoms occupy the Wyckoff $2a(0,0,0)$ 
sites and the four F atoms are located at the $4f(x,x,0)$ sites. Therefore, this structure is characterized by 
three crystallographic parameters: two lattice parameters ($a$ and $c$) and an internal parameter for 
the coordinates of the four F atoms ($x$). The structural parameters obtained using the \pbesol{} functional are $(a,c,x) = (4.7194~{\rm\AA}, 3.1376~{\rm\AA},
0.3037)$, while the corresponding LDA results are 
(4.6373~{\rm\AA}, 3.0990~{\rm\AA}, 0.3033). These calculated values are in very good agreement with the experimental data
(4.7038~{\rm\AA}, 3.1336~{\rm\AA}, 0.3035)\cite{Toole01v57} and (4.7034~{\rm\AA}, 3.1335~{\rm\AA}, 0.303)\cite{Chatterji11v98}.
As expected, the LDA lattice parameters are underestimated while those of the \pbesol{} are slightly overestimated.  

For \bef{}, the three considered crystal structures are $\alpha$-quartz [\Fig\ref{fig:structures}(b)], $\alpha$-cristobalite 
[\Fig\ref{fig:structures}(c)] and $\alpha^{\prime}$-cristobalite. The $\alpha$-quartz phase has trigonal 
symmetry and nine atoms in the conventional hexagonal unit cell. The three Be atoms occupy the Wyckoff $3a(x_1,x_1,0)$ 
sites and the six F atoms occupy the $6c(x_2,y_2,z_2)$ sites. 
On the other hand, the $\alpha$-cristobalite phase has a tetragonal 
symmetry and twelve atoms per primitive unit cell. The four Be atoms occupy the Wyckoff $4a(x_1,x_1,0)$ 
sites and the eight F atoms occupy the $8c(x_2,y_2,z_2)$ sites. 
Therefore, each of these two structures has six crystallographic parameters: two lattice parameters
($a$ and $c$), and four internal parameters (denoted as $x_1$, $x_2$, $y_2$ and $z_2$).
The atomic coordinates of the $\alpha^{\prime}$-cristobalite phase can be obtained 
from those of the $\alpha$-cristobalite by mirror-image transformation $(x, y, z) \rightarrow (-y, -x, z)$, and the 
lattice parameters of the two structures are identical. Therefore, only the structure parameters of the first two crystal 
structures are reported. Our LDA, \pbesol{}, and PBE results, shown in Table~\ref{tab:struct}, are in good agreement with 
available experimental data and other theoretical calculations. The \pbesol{} results lie between the corresponding 
LDA and PBE results and show the best agreement with the experimental data for the $\alpha$-quartz.

\begingroup
\squeezetable
\begin{table}
\caption {\label{tab:elastic} 
Calculated elastic constants of the $\alpha$-quartz and  $\alpha$-cristobalite phases  of \bef{}, and the rutile phase of \znf{}. }
\begin{ruledtabular}
\begin{tabular}{llllllllll}
System & Phase & Approach  & \multicolumn{7}{c}{Elastic constants (GPa) } \\
\cline{4-10} 
               &        &               &  $C_{11}$ & $C_{12}$ & $C_{13}$ & $C_{14}$ & $C_{33}$ & $C_{44}$ &  $C_{66}$ \\ \hline
      \bef{}   & $\alpha$-quartz    &  LDA      & 46.975    & 14.223 & 12.067 &  $-6.401$ & 75.287 & 31.745 &  \\
               &        &  \pbesol{} & 42.278    & 4.991 & 3.760   & $-9.077$  & 53.009 & 30.548 & \\          
               &  $\alpha$-cristobalite      &  LDA      & 33.309    &  7.300 & $-5.087$ &         & 22.487 & 35.810 & 13.731 \\  
               &        &  \pbesol{} & 32.589    &  4.839 & $-5.336$ &         & 24.412 & 37.813 & 16.078 \\  \hline
      \znf{}   & Rutile &  LDA      & 139.442   & 121.550 & 109.127 &       & 220.673 & 36.583 & 91.826 \\
               &        &  \pbesol{} & 128.470   &  98.717& 94.547 &         & 200.902& 35.523 & 83.001 \\
        &    &  Expt.\cite{Rimai77v16}     & 125.5     &  91.8  & 83.0   &         & 192.2  & 39.5   & 80.7   \\   

\end{tabular}
\end{ruledtabular}
\begin{tabbing}
\end{tabbing}
\end{table}
\endgroup

\subsection{Elastic properties and stability}

We show in Table~\ref{tab:struct} the LDA, \pbesol{} and PBE relative energies ($\Delta E$) of the $\alpha$-cristobalite phase with respect to those of $\alpha$-quartz. According to the PBE and \pbesol{} calculations the latter phase is slightly 
more stable, in accordance with the Nelson et al.\cite{Nelson17v95} GGA calculations. However, the LDA calculations lead to an opposite conclusion. Similar conclusions have recently been reached by Masoumi\cite{Masoumi21}, using both LDA and GGA calculations. These results show that these two phases have extremely close cohesive energies.
 
The $\alpha$-quartz phase of \bef{} with a trigonal crystal symmetry has six independent elastic constants\cite{Pabst13v57}.
On the other hand, the rutile phase of \znf{}  and $\alpha$-cristobalite phase of \bef{} (both have a tetragonal crystal
symmetry) also have six independent elastic constants\cite{Pabst13v57}. The elastic constants of these phases, obtained by
using the LDA and \pbesol{} functionals are shown in Table~\ref{tab:elastic}. There are two features to note from this table. 
First, our results for the rutile \znf{} are in very good agreement with the available 
experimental values\cite{Rimai77v16}. Secondly, the \pbesol{} values are systematically smaller than the corresponding LDA values, which is expected since the \pbesol{} GGA functional leads to softer materials than LDA (see above). The calculated elastic constants are used in the calculations of the linear TECs (see Sec.~\ref{sec:method}). 

The elastic constants could be used to investigate the mechanical stability of the crystal structure. For $\alpha$-quartz structure, the Born stability criteria\cite{Pabst13v57} are 
\begin{equation}
 D = (C_{11}+ C_{12})C_{33} - 2C_{13}^2 > 0,   \\ 
\label{eq:K1}
\end{equation}
and 
\begin{equation}
  (C_{11}-C_{12})C_{44} - 2C_{14}^2 > 0.
\label{eq:K2}
\end{equation}
Note that the same expression of $D $ appears in \Eq~\ref{eq:LTEC}.
As for the rutile and $\alpha$-cristobalite phases (of tetragonal (I) class)\cite{Mouhat14v90},  the Born stability criteria are:  
$C_{11} > |C_{12}|$,  $D > 0$, $C_{44} > 0$ and $C_{66} > 0$.
The elastic constants reported in Table~\ref{tab:elastic} show that these criteria are satisfied, 
and therefore the considered phases of \znf{} and \bef{} are mechanically stable. The mechanical stability of these crystal 
structures can also be inferred from phonon dispersion relations, discussed below. 

\begingroup  
\squeezetable
\begin{table}
\caption {\label{tab:DC-EC} 
Calculated high-frequency dielectric constants (DCs) and Born effective charges of the $\alpha$-quartz (AQ) and $\alpha$-cristobalite (AC) phases of \bef{}, and the rutile phase of \znf{}.}
\begin{ruledtabular}
\begin{tabular}{lllcccccccccccc}
\hline
System & Phase & Approach  & \multicolumn{2}{c} {DC} &   & \multicolumn{9}{c} {Born effective charge} \\ 
\cline{4-5} \cline{7-15} 
       &     &           & \it{xx=yy} & \it{zz} &  atom & \it{xx} & \it{xy} & \it{xz} & \it{yx} & \it{yy} & \it{yz} & \it{zx} & \it{zy} & \it{zz} 
\\ \hline \bef{} & AQ & \pbesol{} &  1.902 &   1.912  & Be  &   1.728 &  0.000  & 0.000 &  0.000 &   1.914  &  0.081 & 0.000 &  -0.079 &   1.866 \\
     &    &    &      &      & F   &   $-0.747$ &   0.221 &  $-0.115$ &   0.213 &  $-1.074$ &   0.337 &    $-0.095$  &  0.345  &  $-0.933$ \\
     &      & LDA &   1.887 &  1.896 & Be &  1.727 &  0.000 &  0.000 &   0.000  &  1.918   & 0.080  &  0.000 &  $-0.0758$  &  1.864 \\
      &     &     &      &     &  F & $-0.746$ &   0.226 & $-0.124$  &  0.218  & $-1.076$   & 0.343  & $-0.104$  &  0.356 &  $-0.932$ 
\\ \hline \bef{} & AC   & \pbesol{} &   1.723   &    1.718  & Be &  1.832 &    0.005 &  $-0.045$ &   0.005 &   1.832
     & 0.049  &  0.100 &  $-0.101$ &   1.810 \\
     &  &   &    &     & F  &  $-1.221$  & $-0.117$ &    0.377 &  $-0.101$ &  $-0.611$ &   0.061  &  0.380 &    0.100 &  $-0.905$ \\ 
     &    &   LDA   &   1.827  &   1.818 & Be &   1.823 &   0.005  & $-0.036$  &  0.005  &  1.823   & 0.036  &  0.110 &  $-0.110$  &  1.790 \\
     &        &   &   &     & F  &  $-1.199$ &  $-0.135$ &  0.329 & $-0.120$ &  $-0.624$ & 0.076 &   0.333  &  0.115 &  $-0.896$
\\ \hline \znf{} & Rutile & \pbesol{} & 2.549 & 2.664 & Zn & 2.222  & $-0.162$ &  0.000 & $-0.162$ & 2.222 & 0.00 & 0.000 & 0.000 & 2.424 \\
     &        &          &       &       &  F & $-1.111$ & $-0.409$ &  0.000 & -0.409 & $-1.111$ & 0.000 & 0.000 & 0.000 & $-1.200$ \\
     &        & LDA     & 2.547 &  2.653 & Zn &  2.206 &   $-0.1493$ &  0.000 &  $-0.1493$ & 2.206 & 0.000 & 0.000 & 0.000 & 2.392\\ 
     &        &         &       &        &  F &  $-1.103$& $-0.395$ & 0.000 & $-0.395$ & -1.103 & 0.000 & 0.000 & 0.000 & $-1.19$6\\ 
     &        & Expt.\cite{Alaoui80v41} & 2.6 & 2.1 & & & & & & & & & &  
\end{tabular}  
\end{ruledtabular}
\begin{tabbing}
\end{tabbing}
\end{table}
\endgroup

\begin{figure}[htbp]
    \begin{center}
    \includegraphics[scale = 0.65]{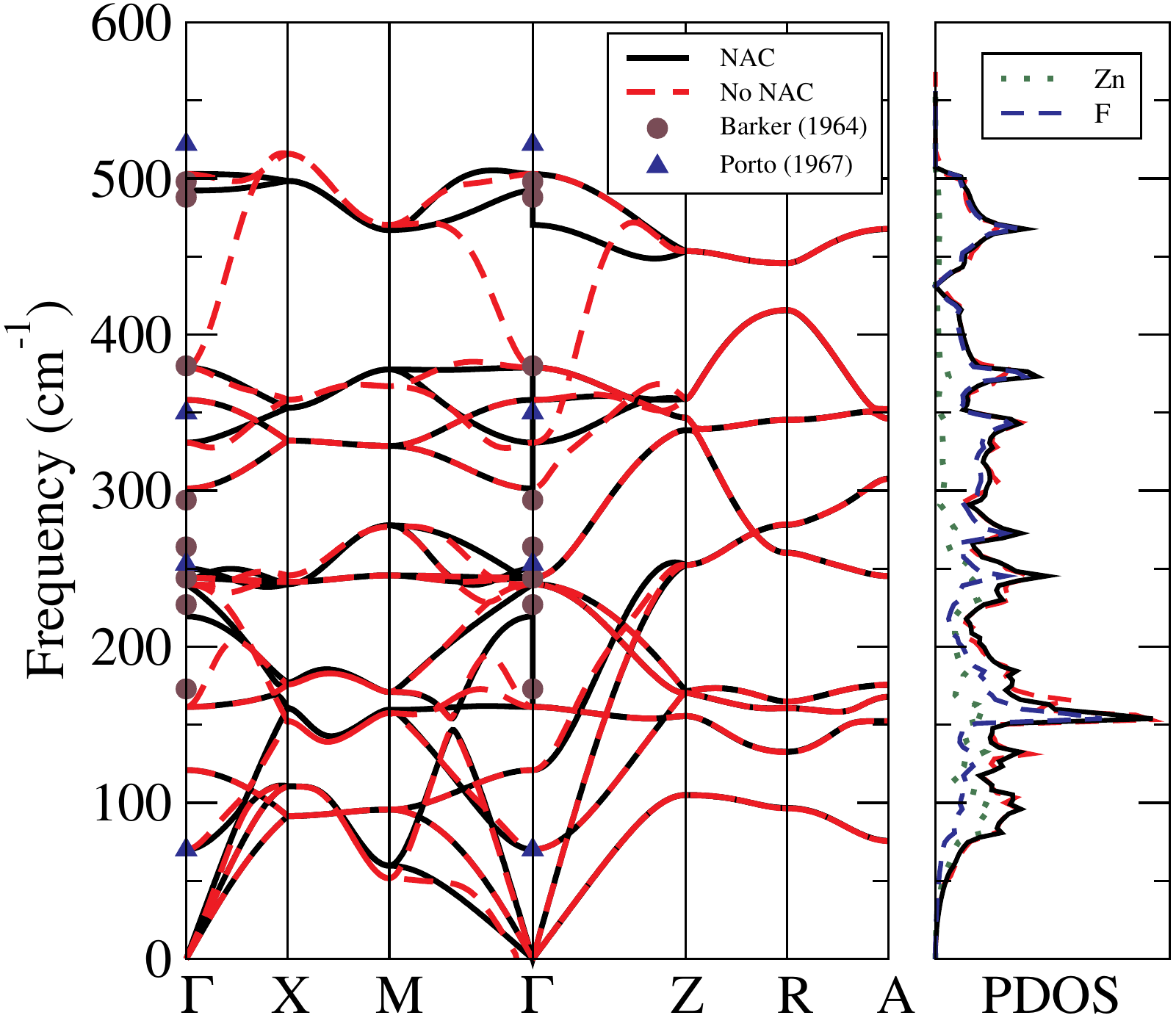} \\ 
    \caption{Calculated phonon dispersion relations and PDOS of \znf, with (black solid curves) and without (red dashed curves)  non-analytic correction (NAC). The Zn and F projected PDOS, with NAC, and also shown. Symbols: available experimental data\cite{Barker64v136,Porto67v154}.}
    \label{znf-pdr}
       \end{center}
\end{figure}

\subsection{Phonon dispersion relations}

Since the considered crystals are polar in character we perform non-analytical correction (NAC) to their dynamical matrices. To do that, we have calculated the high-frequency dielectric constant and Born effective charge tensors, and the results are listed in Table~\ref{tab:DC-EC}. The features to note from this table are the following. 
(i) The calculated results have a very weak dependence on the used exchange-correlation functional. 
(ii) The effective charges in the \znf{} are larger than in \bef{}, which shows that the ionicity of Zn-F bond is larger 
than that in Be-F. This is consistent with the corrected Allred-Rochow electronegativity values\cite{Qteish19v124} 
(larger for Be). 
(iii) Our calculated $xx$ and $yy$ components of the dielectric constant are in very good agreement with available experimental data\cite{Alaoui80v41}, while that of $zz$ is larger than the measured one. However, the comparable values of 
diagonal components of the dielectric constant in our calculations are consistent with the experimental and calculated 
values for other metal fluorides crystallizing in the rutile structure (such as MgF$_2$ and FeF$_2$)\cite{Benoit88v21}.

Fig.~\ref{znf-pdr} shows the calculated phonon dispersion relations and PDOS of the rutile \znf{}, with and without NAC. Also shown are the calculated Zn and F projected PDOS, with NAC, and the available experimental 
data\cite{Barker64v136,Porto67v154}. The frequency spans across an interval of about 500~\invcm.
The features to note from this figure are the following. 
(i) As expected, the NAC leads to longitudinal optical-traverse optical (LO-TO) splitting, near the $\Gamma$ point. The strongest effects are felt by high-frequency optical modes. However, the effects of the NAC on the calculated PDOS are quite small. 
(ii) Experimental data are available only for infrared\cite{Barker64v136} and Raman\cite{Porto67v154,Kurzydlowski20v59} active modes 
at the $\Gamma$-points. The reported frequencies of the latter modes are in very good agreement with each other, and hence only those of Ref.~[\onlinecite{Porto67v154}] are shown in Fig.~\ref{znf-pdr}. For the designation 
of these phonon modes see Ref.[\onlinecite{Benoit88v21}]. Fig~\ref{znf-pdr} shows that these experimental data agree 
reasonably well with our first-principles results.

\begin{figure}[htbp]
    \begin{center}
    \includegraphics[scale = 0.65]{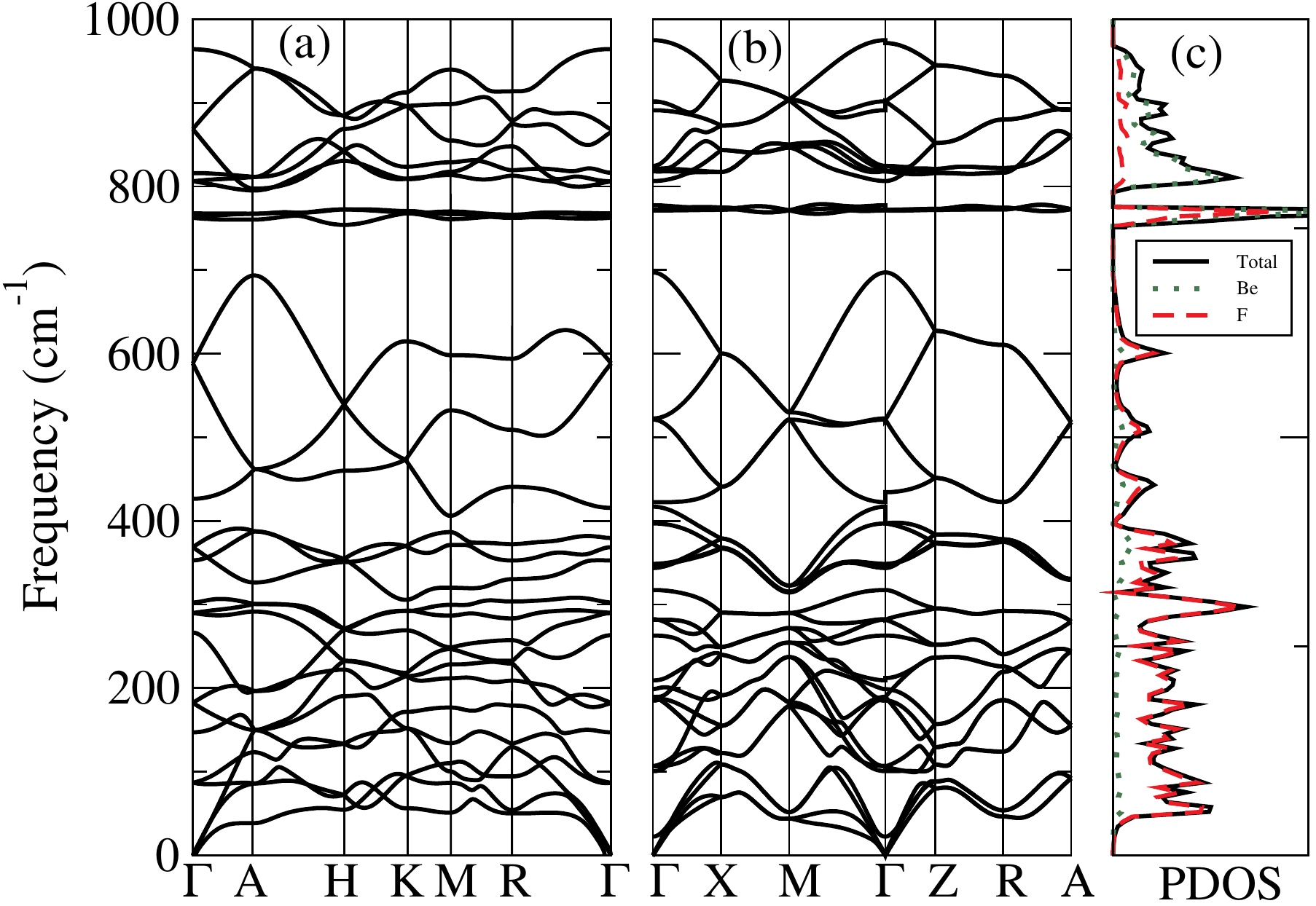} \\
    \caption{Calculated phonon dispersion relations of the (a) $\alpha$-quartz and (b) $\alpha$-cristobalite phases of \bef{}.  (c) PDOS, and the Be and F projected PDOS of the $\alpha$-cristobalite phase.  }
    \label{bef-pdr}
       \end{center}
\end{figure}

\Fig~\ref{bef-pdr} shows the phonon dispersion relations of the $\alpha$-quartz and $\alpha$-cristobalite  phases of \bef{}, taking 
into account the NAC. Also shown are the PDOS, and Be and F projected PDOS of the $\alpha$-cristobalite phase. 
The results of the $\alpha^{\prime}$-cristobalite phase are very similar to those of $\alpha$-cristobalite and hence are not
shown. The features to note from this figure are the following. 
(i) The very wide frequency range of the phonon modes in these systems, compared to that of \znf{}. This can be understood as a consequence of the rather large mass difference between Be and Zn atoms.
(ii) The frequency range of both phases of \bef{} can be separated, according to the character of the phonon modes, into three sub-regions.  (a) The lower frequency region between 0 and about 700~\invcm, where the phonon modes are 
mainly due to the vibrations of F atoms. The contribution of the Be atoms becomes appreciable above 300~\invcm. It is worth 
noting that in the case of \znf{} the dominance of the vibrations of the F atoms occurs in the upper part of the frequency range 
because the Zn atom is heavier than the F atom. (b) A narrow intermediate region at about 770~\invcm, where the rather localized
phonon modes originate from vibrations involving both Be and F atoms. (c) The upper-frequency region, where the phonon 
modes originate mainly from vibrations of Be atoms. (iii) The opening of two frequency gaps, between (a) and (b), and between
(b) and (c) sub-regions. These frequency gaps can be understood as a consequence of the localization of phonon modes in the (b) 
sub-region.

\begin{figure}[htbp]
    \begin{center}
    \includegraphics[scale = 0.65]{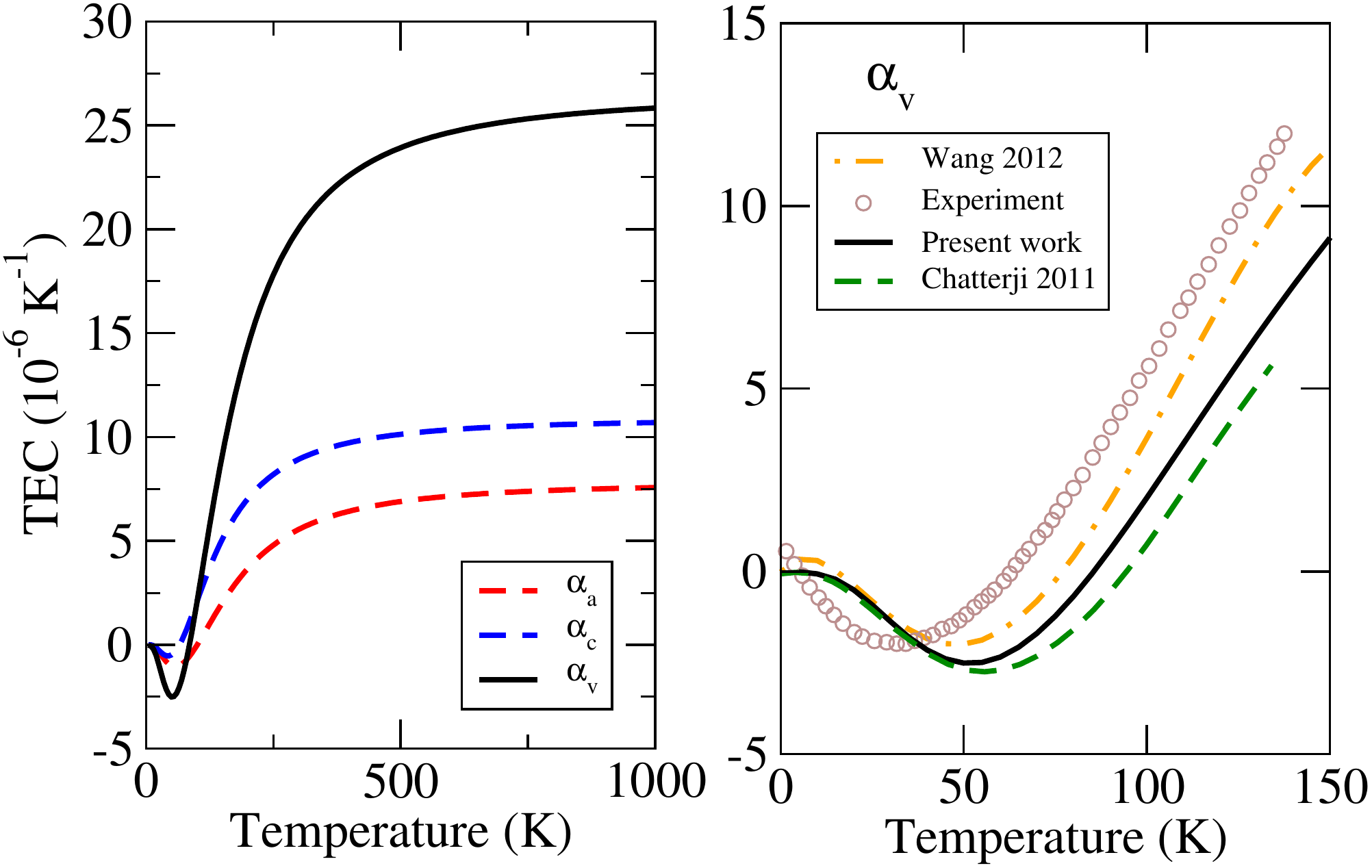} \\
    \caption{Calculated linear and volumetric TECs of \znf{} using the LDA, compared 
             to the available experimental data\cite{Chatterji11v98} and previous theoretical results\cite{Chatterji11v98,Wang12v47}. }
    \label{znf-tec}
       \end{center}
\end{figure}

\subsection{Thermal expansion}
The calculated linear and volumetric TECs of the considered phases of \znf{} and \bef{}, according to the procedure described in Sec.~\ref{sec:method}, are depicted in Figs.~\ref{znf-tec}~and \ref{bef-tec}, respectively. It is worth mentioning that, as expected,
the NACs to the dynamical matrices have negligible effects on the calculated linear TECs.

\begin{figure}[htbp]
    \begin{center}
    \includegraphics[scale = 0.65]{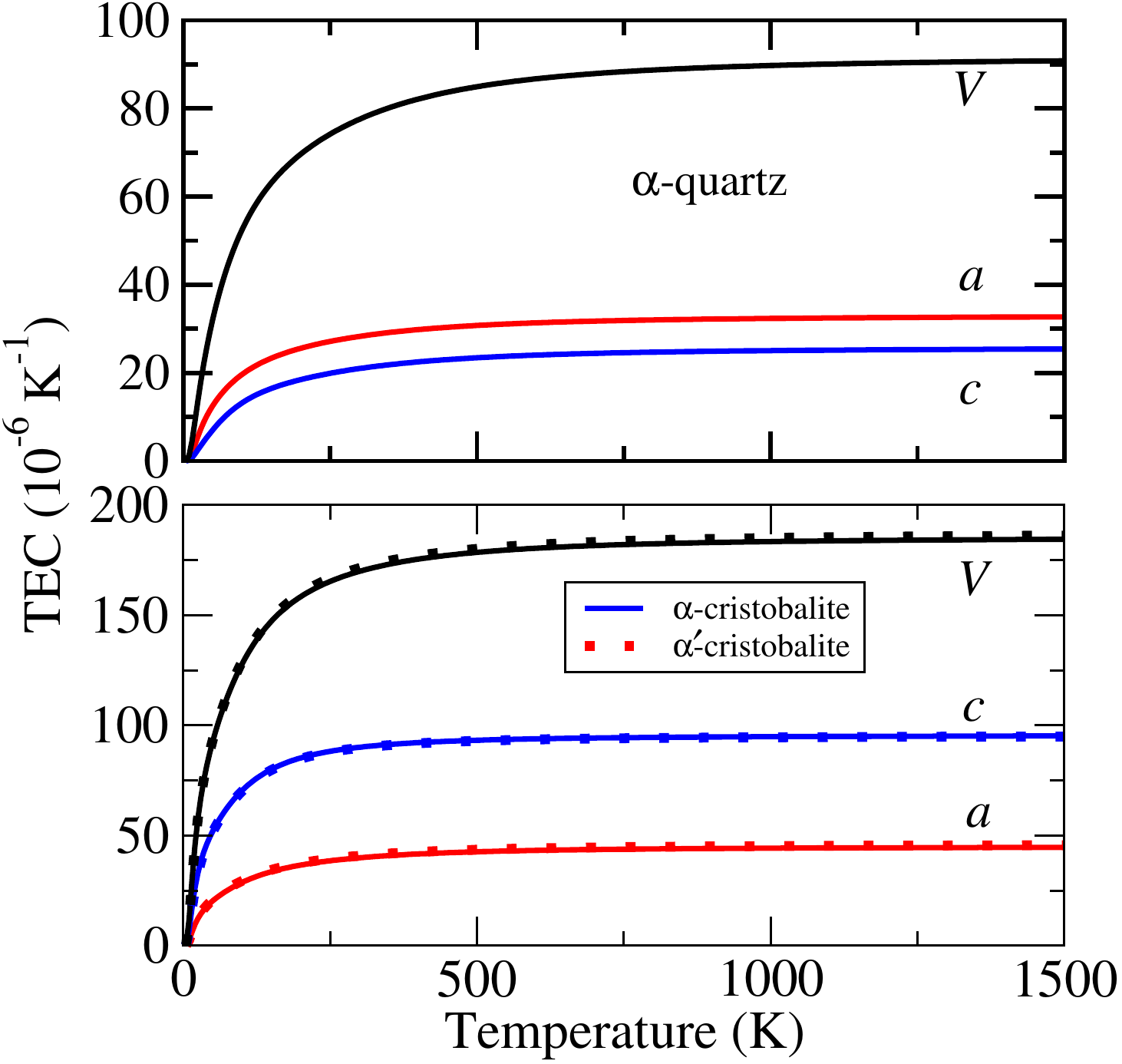} \\
    \caption{Calculated linear and volumetric TECs with \pbesol{} of the three considered phases of \bef{}. Note the difference in the scales of the two panels.}
    \label{bef-tec}
       \end{center}
\end{figure}

We will first look at the TECs of \znf. The important features to note from Fig.~\ref{znf-tec} are the following. 
(i) The NTE at low temperatures is mostly due to $\alpha_a$. The negative values of $\alpha_c$ are smaller (in magnitude) than 
those of $\alpha_a$ and lie in a considerably shorter $T$-range. This is clear from the magnitude and the location
of the minimum values: $ \alpha_a \sim -1.05 \times 10^{-6}$~K$^{-1}$ at $55$~K, and $\alpha_c \sim -0.5 \times 10^{-6}$~K$^{-1}$
at $40$~K. These results are consistent with observed $T$-variations of the $a$ and $c$ lattice parameters at low
temperatures (see Fig.~3 of Ref.~[\onlinecite{Chatterji11v98}]). 
(ii) The calculated $\alphaV$ from the linear TECs (i.e., $\alphaV=2\alpha_a + \alpha_c$) are in good agreement with the previous direct
theoretical calculations\cite{Chatterji11v98,Wang12v47}, and the results of all these theoretical calculations are in a qualitative agreement with experimental data\cite{Chatterji11v98}. This finding reflects the accuracy and reliability of our calculated linear TECs. 
(iii) $\alpha_c$ is systematically and appreciably larger than $\alpha_a$. For example, at $300$~K the calculated value of $\alpha_c$ 
(of $8.9 \times 10^{-6}$~K$^{-1}$) is about $60$\% larger than that of $\alpha_a$ (of $5.6 \times 10^{-6}$~K$^{-1}$).

As for the thermal expansion of the considered phases of \bef, the features to note from Fig.~\ref{bef-tec} are the following. 
(i) Unlike \znf, the calculated values of both $\alpha_a$ and $\alpha_c$ are always positive for all of the considered phases of \bef.
(ii) Both $\alpha_c$ and $\alpha_a$ of the $\alpha$-cristobalite structure are very close to those of $\alpha^{\prime}$-cristobalite, and hence only those of the former phase will be discussed below.
(iii) In the considered $T$-range, $\alpha_a(T)$ of the $\alpha$-quartz structure is slightly larger than $\alpha_c(T)$, whereas $\alpha_a(T)$ of the $\alpha$-cristobalite phase is much smaller than $\alpha_c(T)$.   
(iv) The large $\alpha_c$ and $\alpha_a$ lead to very large $\alphaV$ for both phases of \bef{}. For example, at $300$~K,
the values of $\alphaV$ are of $77.6$ and $169.9$ $\times 10^{-6}$~K$^{-1}$
respectively for the above two phases of \bef{}, compared to that of $20.0 \times 10^{-6}$~K$^{-1}$ for \znf{}.  
Our largest calculated linear TEC is of $\sim 95 \times 10^{-6}$~K$^{-1}$ at $300$~K for $\alpha_c$ of the $\alpha$-cristobalite phase. 
This is indeed large compared to the experimental linear TECs at $300$~K of four fluorites, i.e., CaF$_2$, SrF$_2$, BaF$_2$, and PbF$_2$\cite{Roberts86v19} that range between $18.1$ and $29~\times 10^{-6}$~K$^{-1}$, but still is somewhat
smaller than the measured linear TEC value of $163.9 \times 10^{-6}$~K$^{-1}$ of an Ti-Nb alloy\cite{Bonish17v8}.

\begin{figure}[htbp] 
    \includegraphics[scale = 0.65]{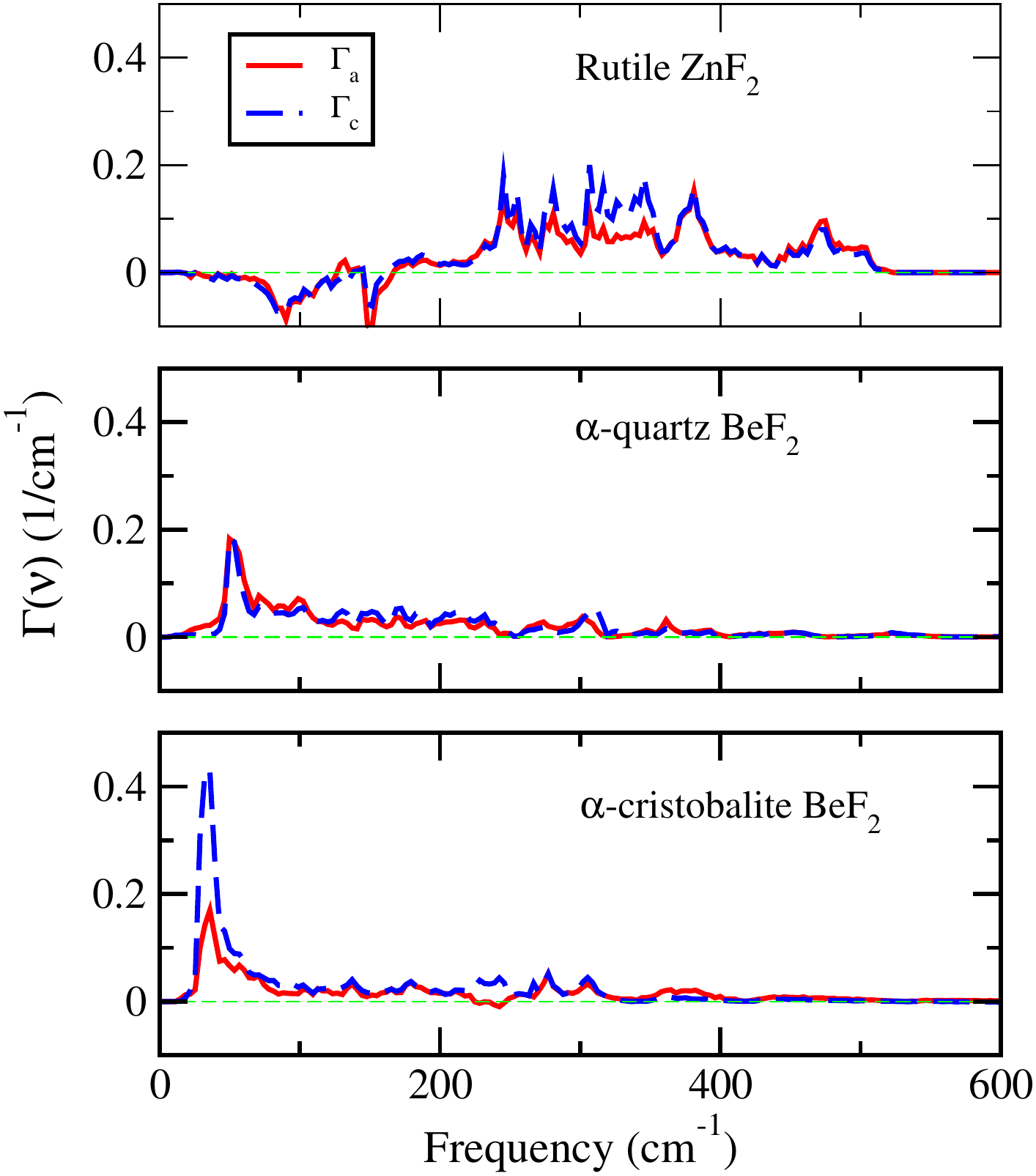} \\
    \caption{The PDOS weighted by the \gru{} parameters of the rutile \znf{}, and the $\alpha$-quartz and $\alpha$-cristobalite phases of \bef{}.}
    \label{fig:gammanu}
\end{figure}

The key physically insightful quantity for the interpretation of the above results is the PDOS weighted by 
the \gru{} parameters, $\Gamma_i(\nu)$, defined in \Eq~\ref{eq:Gammanu}.
\Fig~\ref{fig:gammanu} shows $\Gamma_i(\nu)$ of \znf{}, and the $\alpha$-quartz and $\alpha$-cristobalite phases of \bef{}. 
The important features to note from this figure are the following. 
(i) For \znf{}, the low-frequency modes ($\nu < 150$~\invcm{}) have negative \gru{} parameters, which lead to negative 
$\Gamma_i(\nu)$ in this $\nu$ range. Since low-frequency modes are easily thermally excited, this finding explains the observed NTE in \znf. Moreover, by inspecting the differences between $\Gamma_a(\nu)$ and $\Gamma_c(\nu)$ one can easily understand why $\alpha_a$ is always lower than $\alpha_c$.
(ii) The $\Gamma_i(\nu)$ of the considered phases of \bef{} are always positive, which reflects the dominance of 
positive mode \gru{} parameters in these phases. This explains the lack of NTE in the considered phases of \bef. 
(iii) The $\Gamma_a(\nu)$ and $\Gamma_c(\nu)$ of $\alpha$-quartz \bef{} have comparable magnitudes, with $\Gamma_c(\nu)$ 
being smaller than $\Gamma_a(\nu)$ for $\nu < 100$ \invcm{}, which explain the comparable magnitudes and ordering of 
its $\alpha_c$ and $\alpha_a$.
(iv) The peak in $\Gamma_c(\nu)$ of the $\alpha$-cristobalite phase around $\nu \sim 34$~\invcm{} is much higher than that 
of the $\Gamma_a(\nu)$, which results in a large $\alpha_c$ compared to $\alpha_a$. This finding means that, in this $\nu$-range, 
the positive mode \gru{} parameters associated with the out-of-plane deformation are significantly larger than those associated 
with the in-plane deformation. The large \gru{} parameters can be viewed as a manifestation of strong anharmonic effects in the $\alpha$-cristobalite and $\alpha^{\prime}$-cristobalite structures of \bef{}. However, it should be noted that large \gru{} parameters 
are not the only factor that is responsible for the giant $\alphaV$ of the $\alpha$-cristobalite: the elastic property
via the negative (and with a large magnitude) $C_{13}$ elastic constant 
(see Table~\ref{tab:elastic} and \Eq\ref{eq:LTEC}) plays also a major role.
The above findings explain the much larger volumetric TEC of the $\alpha$-cristobalite phase of \bef, compared to that of the 
$\alpha$-quartz phase, which, in turn, is larger than that of \znf{}.

\section{Summary}
First-principles calculations are performed to investigate the structural, elastic, and vibrational properties of the 
rutile structure of \znf{} and three crystal structures of \bef{} ($\alpha$-quartz, $\alpha$-cristobalite and its similar phase 
with space group \PFourThree{}). The so-obtained phonon density of states, mode \gru{} parameters, and elastic constants are used to study the linear thermal expansion coefficients (TECs) of the compounds mentioned above,
within a \gru{} formalism. We have used deformations that preserve the symmetry of the crystal to obtain the \gru{} parameters. The considered crystal structures of both 
systems are found to be mechanically stable. The calculated physical quantities for 
both systems are in very good agreement with the available experimental data and previous theoretical results. For 
\znf{}, the calculated linear TECs, $\alpha_a$ and $\alpha_c$, along the $a$ and $c$ directions are consistent with 
the experimental $T$-variations of the corresponding lattice parameters, respectively. The volumetric TEC
$\alphaV$ computed from these linear TECs is in qualitative agreement with experiment at low temperatures, including 
negative thermal expansion (NTE) behavior. The considered phases of \bef{} are not NTE materials, and their linear TECs
are much higher than those of \znf{}, especially for the $\alpha$-cristobalite phase. The elastic constants,
high-frequency dielectric constants, Born effective charge tensors, and TECs of the considered phases of \bef{} are
reported in this work for the first time and could serve as predictions.

\section{Acknowledgment}
We thank the National Supercomputing Center, Singapore (NSCC) and A*STAR Computational Resource Center, 
Singapore (ACRC) for computing resources. We also thank Iyad Al-Qasir for fruitful 
discussions. This work is supported by RIE2020 Advanced Manufacturing and Engineering (AME) Programmatic Grant No A1898b0043.

\bibliography{master-references}

\end{document}